# Evaluating and improving social awareness of energy communities through semantic network analysis of online news

Piselli, C., Fronzetti Colladon, A., Segneri, L., & Pisello, A.L.

# Evaluating and improving social awareness of energy communities through semantic network analysis of online news

**Abstract**

The implementation of energy communities represents a cross-disciplinary phenomenon that has the potential to support the energy transition while fostering citizens' participation throughout the energy system and their exploitation of renewables. An important role is played by online information sources in engaging people in this process and increasing their awareness of associated benefits. In this view, this work analyses online news data on energy communities to understand people's awareness and the media importance of this topic. We use the Semantic Brand Score (SBS) indicator as an innovative measure of semantic importance, combining social network analysis and text mining methods. Results show different importance trends for energy communities and other energy and society-related topics, also allowing the identification of their connections. Our approach gives evidence to information gaps and possible actions that could be taken to promote a low-carbon energy transition.

**Highlights**

- Italian online news is analyzed through semantic network analysis and text mining.
- We use the Semantic Brand Score to measure the media importance of energy communities.
- We identify insights for implementation and societal awareness of energy communities.
- The media importance of energy communities is lower than other energy-related topics.

## 1. Introduction

The European Green Deal presents the European Union (EU) strategy for driving the energy transition to achieve a better society through a resource-efficient and competitive economy [1]. To this aim, the energy transition is driven toward reducing the use of non-renewable natural resources and achieving zero net greenhouse gas emissions by 2050 [2]. Among the strategies involved on the road to this low-carbon energy transition, energy communities are a promising solution [3]. Two main types of energy communities are considered at the European level: Renewable Energy Community (REC) and Citizens Energy Community (CEC), introduced in the recast of the Renewable Energy Directive (RED II) 2018/2001/EU [4] and in the Internal Electricity Market Directive 2019/944/EU [5]. These directives entered into force as part of the Clean Energy for all Europeans Package [6]. Energy communities are defined as legal entities involving citizens' participation as prosumers in the future energy system that should integrate social justice principles. The main goal of these legal, social, and technical entities is to provide environmental, economic, and social benefits for the community rather than profit-making [7]. They can be organized in various collective forms for the decentralization and the local operation of renewable energy [8]. Energy communities have already been demonstrated to be an effective strategy for changing the energy system in some early adopting countries [9], with a focus of some countries on renewable energy [10]. However, their spread depends on providing a consistent and favorable regulatory framework in the different European countries, ensuring the cost-effectiveness of consumer-funded efficiency investments, and leveraging effective consumer engagement strategies [11].

In this panorama, Italy, which acknowledges only renewable energy communities, falls back in the adoption of the above-mentioned strategy. The "energy community" concept was firstly introduced on a national basis by the "Milleproroghe" decree n. 162, of December 30, 2019 [12], according to the Directive 2018/2001/EU. In other words, the law allowing the establishment of energy communities was only recently issued (law n. 8, of February 28, 2020) [13]. However, the regulatory framework is still under development and continuously changing, with new policies that share the goal of reducing GHG emissions. Two similar schemes of energy community are feasible in Italy: renewable energy communities and collective self-consumption schemes. Both are based on renewable energy sharing among community members, with specific constraints in terms of tariffs, technologies, and the amount of shared energy. The specificity of collective self-consumption schemes is that they are established among subjects in the same building or condominium. At the same time, Italian energy communities were meant to install a relatively small amount of shared energy (less than 100 kW), imposing a granular size of those installations. However, these boundaries are still under discussion for a possible increase in size and diffusion, opening the door to more significant deployment of renewables, even other than the classic photovoltaic systems [14].

In addition, although renewable energy communities are starting to be also developed in laggard countries (such as Italy), there seems to be a gap between technological development – which is extremely advanced in terms of renewable solutions and platform management and control systems [15] – and citizens' understanding and engagement towards their deployment that worries policymakers [14]. In particular, this societal awareness gap is extended to professionals who design large-scale projects and represent key stakeholders. In this context, Lagendijk et al. [16] showed that technological progress and economic viability aspects seem more powerful than environmental and social triggers. In early-adopter European countries, e.g., the Netherlands or Germany, the growth of RECs is somehow declining, as pointed out by Herbes et al. [17]. They showed that outdated business models could be one obstacle to implementing RECs. Another major problem is often the lack of and termination of supportive economic policy instruments, such as feed-in-tariffs [18,19]. Traditionally, cooperatives are the most diffused [20], where citizens collectively own and manage renewable energy projects. Indeed, the social structure behind energy communities is particularly relevant for their diffusion [21]. Prospective users' awareness and acceptance represent the fundamental basis for effectively exploiting innovative technologies [22]. This approach involves both people's engagement in the energy system and changing the system to consider the needs of people [23]. For a larger diffusion of energy communities, people have to be prepared to engage in new roles, e.g., prosumers, by implementing bottom-up processes to meet energy agents', policymakers', and consumers' expectations [24].

In this paper, we analyze online news, following the idea that online media can be leveraged for citizens' engagement [25]. Online news, in particular, allow mass communication. They reach many ordinary citizens who could be the target of energy community communication campaigns [26]. Online news can also be regarded as a proxy of the media message conveyed on other channels (such as printed news or TV broadcasts). In general, for a successful diffusion and implementation of energy communities, not only energy providers and field experts should be reached by energy transition-related knowledge. Direct citizens' involvement is pivotal, as they are key decision-makers, as also confirmed by the last report on the energy efficiency of the International Energy Agency (IEA) [27]. In section 2, we further comment on the role of online media in pushing the energy transition.

Considering this scenario, the goal of this work is to present an analysis of the media importance of energy communities and their related topics. We used a methodology that combines text mining and social network analysis to achieve this goal, for the first time applied to the discourse around energy communities. In particular, we present a new approach that can be used to automatically analyze online news to evaluate the importance trends of energy-related topics (including energy communities) and their textual association. This approach allows a continuous monitoring of time trends and topics to get potentially valuable insights for policymakers and practitioners interested in the societal uptake of green energy solutions. For example, it shows which green-energy solutions should be further debated because they are not well-covered – while also giving evidence of their associations with specific themes or topics of high interest. In addition, our analysis shows which events generate higher media buzz and, therefore, could be used to foster citizens' awareness of energy communities.

The remainder of this paper is organized as follows. In the next section, we discuss the importance of media in pushing the energy transition and in analyzing news data to evaluate media importance as a proxy for societal awareness and acceptance of energy technologies and, in particular, energy communities. Section 3 presents the data collection and methodology used. Section 4 describes the main results of the analysis. Section 5 discusses the outcomes, limitations, and future developments of this work.

## 2. The role of media in fostering energy transition

Researches have been studying the societal awareness and acceptance of innovative technologies in the field of energy transition [28]. Frequently, users' acceptance analyses are carried out by means of dedicated interviews or surveys [29]. If extended surveys are submitted within a broad context, researchers can achieve representative and generalizable results [30,31]. For instance, regarding the social acceptance of transitions towards local energy communities, Azarova et al. [32] showed that in European countries such as Italy and Switzerland, a key role is played by national and regional political support. However, surveys and interviews involve a heavy effort to be submitted to a representative number of respondents and may require external support [31]. Moreover, given the subjective nature of survey results, they can be biased due to respondents' or survey characteristics, e.g., the high technical level of the topics addressed, such as often happens in the case of energy-related discussions [33].

While the study of online news might not be enough for a complete assessment of societal awareness and acceptance of energy communities [16], they still represent an important data source that can be analyzed to provide a general view [34]. Indeed, news articles can strongly impact public sentiment [35] and behavior [36], and media coverage is often a proxy for societal awareness of a topic. In addition, online news is widely available and easily accessible to the general public at little or no cost. In general, the choice of focusing on online news is supported by past research that considered news, or web coverage [37], as a potential enhancing factor for the diffusion of green technologies – looking, for example, at biogas production in Finland [38], renewable energy in Canada [39], or the energy scenario in Germany [40]. Thanks to the relatively fast analysis of big textual data, these approaches allow the description of complex relationships and provide insights in almost real-time. For instance, both Lyytimäki et al. [38] and Scheer [40] showed that citizens' perceptions and expectations about energy technologies and strategies depend on the media's strategic information flow. Similarly, Lyytimäki [41] stressed the importance of mass media coverage in implementing climate policies. The author found a wide permeation of climate-change-related issues in different fields of newspaper coverage, which could sustain the spread of this topic into broader policy domains. Pelkonen and Tapaninen [42] explored news media data to study trends of diffusion of renewable energy sources and confirmed the ability of this approach to reflect the social discussion.

Semantic network analysis proves to be a promising approach for analyzing online news [43], providing information from the study of the co-occurrence of concepts and their semantic associations through automatic coding methodologies [44]. It can reveal essential relationships among concepts in the energy domain. For example, it has been applied to investigate the concept of social sustainability and its link with sustainable development [45]. Bickel confirmed the suitability of this approach to assess semantic sustainability by stressing the need to improve sustainability communication and education towards the energy transition [46]. Fariña García et al. [44] investigated the connections among Sustainable Development Goals via semantic network analysis. They found that some goals, including sustainable cities and communities, are better connected. In contrast, others, including those on affordable and clean energy and climate action, are more independent.

With respect to the existing studies, we use a new approach to analyze news and its semantic networks, e.g. [40,43,44]. To the best of our knowledge, we are the first to use this specific approach to study the media attention on the topic of energy communities. There are other studies that have evaluated the media discourse about different energy-related topics – such as renewable energy [37,39], biogas [38], the energy system [40], climate change [41] and the related awareness [47], renewable energy technologies diffusion [42], and solar energy [43]. To the best of our knowledge, only Lagendijk et al. [16] have examined the media framing of (Dutch) energy community initiatives to evaluate their social acceptance. However, none of the above-mentioned studies has used semantic network analysis, also focusing on the interrelationships between energy communities and other topics of the energy discourse to understand their media importance. Most of these studies used a methodology based on quantitative content analysis, and only a few [44–46] used a semantic network approach. For example, Ludovico and colleagues [48,49] adopted a mixed-method approach where they combined social network analysis and computational textual analysis. However, they studied hyperlink networks – with nodes representing actors engaged in the energy transition – or used semantic networks for topic modeling or keyword extraction. Here we adopt a different methodology to evaluate the media importance of energy communities and other energy-related topics, as detailed in section 3. We go beyond the analysis of word frequencies and consider the uniqueness, heterogeneity, and connecting power of semantic associations to better evaluate the potential impact of news content on citizens' memory and behavior [50], as also illustrated in the next section.

### 2.1. The influence of news on people's behavior

Media priming studies support the idea that news can impact people's behavior. Priming procedures first appeared in cognitive psychology, i.e., the science branch studying the structure and representation of information within memory [51]. According to psychological network models of memory, information is stored in memory in the form of nodes, where each node represents a concept. These nodes are connected to related nodes via associative pathways, and the distance between them indicates their link's strength. A further supposition is that each node has an activation threshold; if the activation level exceeds its threshold, the node fires. When a node is activated, for example, when the image of plastic bottles on the beach activates "environmental pollution" or the concept of mitigation actives "climate change", this activation can propagate to other related nodes, for example, "concern". One consequence of propagating activation is that the corresponding node now needs less additional activation to fire; in other words, the activation of nodes increases their accessibility in memory, so they are "primed" by other stimuli [50]. In brief, the primary media-priming process consists of two phases. In the first one, information received through a media channel (in our work, online news data) activates preexisting associated knowledge in the mind of the reader (i.e., "available" cognitive units). The cognitive unit's accessibility is increased by this activation, which implies that the reader will probably use them in interpreting and evaluating a subsequent target stimulus (i.e., the attitude object). If the reader applies the primed concept to a target stimulus when they would not otherwise have done this, then the media priming effect occurs. Therefore, the second phase speaks to the consequences of the priming process [52]. Media priming is "a mnemonic model of information-processing, which assumes that individuals form attitudes on the strength of considerations that are most salient, and thus most accessible when making decisions" [52]. For these reasons, we decided to analyze Italian online news to understand how important the topic of energy communities is and evaluate its connections and potential social awareness. Consistently, we used a measure of semantic importance that not only considers media coverage but also assesses the power and uniqueness of energy communities' associations with other discourse themes.

Our intuition is also supported by past studies showing that news can be a good predictor of social, economic, and political phenomena [53–55]. For example, Park et al. [56] tested the role of news discourse in forecasting tourist arrivals in Hong Kong and confirmed that including news data significantly improves forecasting performance. Fronzetti Colladon [54] used news data to make political forecasts and show the impact of news on election outcomes. Similarly, Lerman et al. [35] discussed the impact of news on public perception of political candidates. Other researchers used sentiment and discourse analysis to assess the general mood towards sensitive issues, such as domestic violence and depression, during the COVID-19 pandemic [57,58]. Online news was also used to understand and increase awareness of specific topics. Boulianne [25] examined how online news affects civic awareness and civic and political life engagement. Their findings suggest that online news increases civic awareness, indirectly affecting engagement. Gil de Zúñiga et al. [59] showed that "seeking information via social network sites is a positive and significant predictor of people's social capital and civic and political participatory behaviors, online and offline." Online news was also shown to significantly affect people's awareness of delicate environmental issues. By analyzing the content of American news, Bolsen and Shapiro [60] investigated how people understand climate change and the actions they are ultimately willing to support to address the problem. Duffy [61] used lay press narratives about how youth respond to climate change to examine the dynamics between minors and adults around the evolving climate crisis. Finally, other relevant studies highlighted the crucial role played by news in important environmental issues – such as the protection of coral reefs [62].

The purpose of this work is to investigate the societal awareness on energy communities in Italy through the analysis of online news to evaluate the media discourse and identify barriers and opportunities for their diffusion and social acceptance. This study looks at energy communities from a new perspective by addressing their societal uptake, understanding, and exploitation through a novel semantic approach - which has the potential to complement the information provided by more traditional approaches, such as surveys and interviews.

## 3. Methods

To evaluate the media importance of energy communities in Italian online news, we used the Semantic Brand Score (SBS) [63]. This composite indicator includes the assessment of three dimensions: (i) *prevalence*, a proxy for the media coverage of a specific topic, usually identified by a set of keywords that we call *energy communities*-related keywords (ECKs) in this work; (ii) *diversity*, i.e., the heterogeneity and uniqueness of the words associated with those of interest; and (iii) *connectivity*, i.e., the ability of ECKs to link different concepts, words, or discourse topics. This measure of semantic importance is recent but not entirely new since it has already been conceptualized, used, and validated in previous studies [53,54,63]. However, to the best of our knowledge, this is the first time it has been applied to the energy-community field of research. The implemented research method follows three steps described in detail in the following sub-sections. Firstly, we collected news from a database containing articles published in the major Italian online news outlets (see Section 3.1). Subsequently, we defined ECK clusters using a multiple-step approach (Section 3.2). Lastly, we calculated their semantic importance through the SBS indicator and studied their textual associations (Section 3.3).

### 3.1. News data collection

News data were collected and made available by Telpress International[1], which has an extensive database containing articles from all major Italian online news sources, covering all topics. In this study, we analyzed about 143,000 online news related to energy topics and published from January 2017 to October 2020. Our analysis focused on the Italian context because Italy is one of the European countries that fall back in implementing energy communities and is currently transposing Directive 2018/2001/EU. This gap among European countries is addressed by various European Actions (Horizon 2020 projects and others) [64], which promote RECs implementation taking early adopter nations, such as The Netherlands, as a virtuous example. Accordingly, all the collected articles were written in the Italian language.

### 3.2. Definition of ECKs

The ECK clusters were defined through a multiple-step approach. Firstly, we used the SBS BI app to extract the most relevant keywords from the news, using a TD-IDF logic[2] [65]. In particular, we identified words that had a high frequency but did not appear in all or most of the text documents (which is typical of stop-words). Subsequently, we manually organized the keywords into groups associated with specific topics, extending the original set based on our knowledge. Eight ECKs emerged from the analysis as the most relevant to the topics of energy communities and energy transition: *energy community*; *collective self-consumption*; *efficiency*; *sustainability*; *renewables*; *social community*; *tax benefits*; *EU directives*, as detailed in Table 1. Some of these ECKs have a broader scope, while others address specific themes. However, the primary purpose of our analysis is not to compare their media importance but to assess the media importance of *energy community* and its relationship with the other topics.

**Table 1**. ECKs clusters and their definition.

| ECKs cluster | Definition |
| --- | --- |
| *Energy community* | Words related to energy communities and peer-to-peer energy trading, including renewable energy sharing and the related strategies. |
| *Collective self-consumption* | Words related to the self-consumption and production of renewable energy, especially collectively. |
| *Efficiency* | Words related to the efficiency and performance of buildings and energy technologies. |
| *Sustainability* | Words related to environmental and energy sustainability, environmental impacts, and greenhouse gas emissions. |
| *Renewables* | General words related to renewable energy sources. |
| *Social community* | Words related to the social valence of communities, collectivity, and social participation. |
| *Tax benefits* | Words related to different types of economic benefits and incentives related to buildings and energy aspects. |

---

[1] http://www.telpress.com/
[2] We also tested other approaches for determining the initial set of keywords – such as the TextRank algorithm [80] without obtaining significantly different results.

| *EU directives* | Reference to relevant EU directives and their reference number, e.g., 2018/2001/EU and 2019/944/EU. |
|---|---|

### 3.3. Measuring semantic importance

Our evaluation of media importance of energy communities and energy-related topics was carried out using the Semantic Brand Score indicator [63] and the SBS Brand Intelligence (BI) web app[3] [66] – using the computing resources of the ENEA/CRESCO infrastructure [67]. This approach, which combines social network analysis and text mining methods, allows automating the analysis of textual data, reducing its complexity and extending its informative power. While some of the studies mentioned in Section 2 apply a semantic network approach for analyzing the discourse about energy communities and other topics, they mostly rely on the count of word frequencies, thus neglecting additional important information captured by the SBS dimensions of diversity and connectivity. We analyze online news to produce a social network where nodes are words/concepts that appear in the articles while links represent their textual relationships.

In particular, we use the Semantic Brand Score (SBS), a measure of semantic importance applicable to (big) textual data. It can be calculated for any word or topic in corpora, not necessarily a "brand". In this work, we use it to evaluate the media importance of a set of pre-defined keywords related to energy communities and some energy transition topics. Building on the information provided by the SBS, we evaluate the relevance of ECKs based not only on how often they are mentioned – as done by other studies – but also on the variety and uniqueness of their textual associations and their connective power in the discourse. Indeed, the SBS is composed of three dimensions: (i) prevalence, (ii) diversity, and (iii) connectivity, which are described in the work of Fronzetti Colladon [63] and briefly recalled in this section. Prevalence measures the frequency of use of a specific word in a set of text documents. It can be regarded as a proxy for awareness and media coverage of a concept because the more an ECK is repeated, the more it can be remembered and recognized. Diversity measures the heterogeneity and uniqueness of the words associated with an ECKs cluster, drawing from the idea that diverse and unique lexical associations can increase an ECK awareness [68,69], which could translate into stronger media priming and memorability of green-energy solutions [50]. Connectivity represents the cluster's ability to bridge connections between other words or topics. This is another dimension related to the strategic value of textual associations, which is similarly important to increase topic memorability. It was also regarded as a proxy of topic popularity in past research [70]. The sum of these three standardized dimensions measures ECK importance according to Eq. (1).

$$SBS(x_i) = \frac{P(x_i)-\bar{P}}{std(P)} + \frac{D(x_i)-\bar{D}}{std(D)} + \frac{C(x_i)-\bar{C}}{std(C)} \qquad (1)$$

Where $x_i$ denotes an ECK, $P(x_i)$ is its prevalence, $D(x_i)$ is its diversity, and $C(x_i)$ is its connectivity. Mean values of the three dimensions ($\bar{P}, \bar{D}, \bar{C}$) and standard deviations ($std$) are

---
[3] https://bi.semanticbrandscore.com

calculated on the sample of ECKs chosen for analysis (for each timeframe). Operatively, to calculate the SBS, the textual data must be pre-processed. Indeed, while prevalence, referring to the presence of a specific ECK in the text, is calculated as a frequency count, diversity and connectivity derive from centrality metrics of social network analysis [71] and require a preliminary transformation of the text corpora into a network of co-occurring words. Text pre-processing procedures aim to reduce the language complexity and retain the most important words [72]. The main steps at this stage are the removal of punctuation, stop-words, and special characters; the transformation of the whole text into lower case; the extraction of stems by removing the affixes of words [73].To give an example, the sentence "*Community energy initiatives are offering new opportunities for citizens to get actively involved in energy matters.*" would be transformed into a list of words as the following (stemming is avoided for the sake of reliability):

[community, energy, initiatives, offering, new, opportunities, citizens, get, actively, involved, energy, matters].

After the pre-processing phase, documents are transformed into undirected networks based on word co-occurrences. Figure 1 shows an example of the network generated if the above-mentioned sentence is considered. This network comprises 11 nodes– each representing a word – and 30 arcs (considering a co-occurrence range of three words as an example). This specific data transformation process allows the calculation of diversity and connectivity. In particular, diversity is operationalized through the distinctiveness centrality metric [74], which penalizes connections to overly connected words. Connectivity is measured using weighted betweenness centrality [75], which counts how often an ECK lies in the shortest network paths that interconnect all the node pairs. This measure can be intended as a proxy of the brokerage power of a keyword, i.e., its ability to connect different parts of the overall discourse [63].

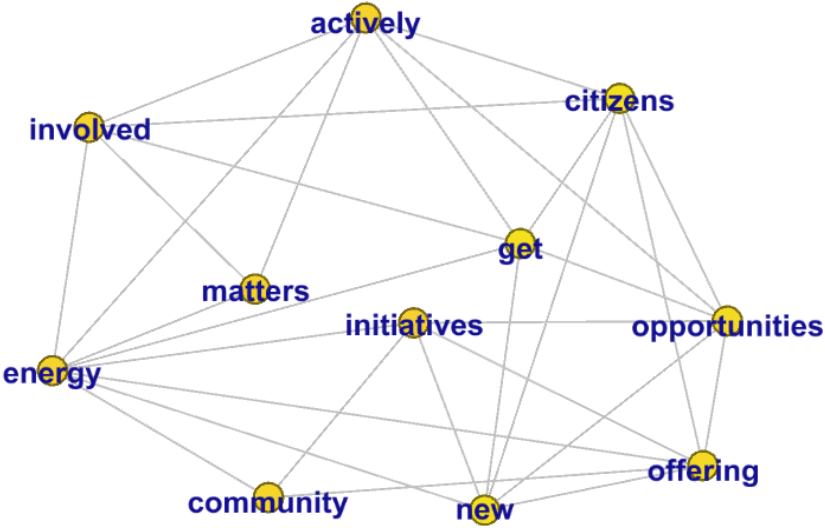

**Figure 1**. Example of co-occurrence network.

As an additional step of our analysis, we also calculated the image similarity of ECK clusters. In particular, we wanted to understand if their textual associations were similar – i.e., how close was the discourse around each cluster and if they were associated with similar concepts/topics. We used a bag-of-words approach, preserving multiplicity, and constructed a document per term matrix, where each row represented a cluster while columns represented their associations. Matrix cells were populated with association frequencies. Subsequently, we calculated an ECK per ECK distance matrix using the cosine similarity metric typically employed in text mining [76]. Similarities were plotted in two dimensions using the multidimensional scaling technique [77]. Results are presented in Section 4.

## 4. Results

We present the results of our analysis in the following, starting with the discussion of ECKs' importance time trends. Figure 2a reports the monthly values of the SBS index for all the eight ECKs. It shows how *social community* and *sustainability* have importance values much higher than all the other ECKs, during the whole period – meaning that these two topics have been largely debated in online news in the last years. Moreover, the monthly trend of the two ECKs is rather similar, except for two peaks of *social community* at the beginning and at the end of the analyzed period. This result suggests that the themes of *social communities* and *sustainability* are always related to each other. They are followed per importance by *renewables, efficiency, energy community, and tax benefits*, which, however, have much lower SBS values. It is worth noting that the SBS values for these ECKs have different, less regular, fluctuations. Lastly, the trend of *collective self-consumption* and *EU directives* is relatively flat and close to zero over the whole period. This result shows that these two ECKs are not particularly debated in online news, leading to reduced diffusion and understanding among Italian stakeholders.

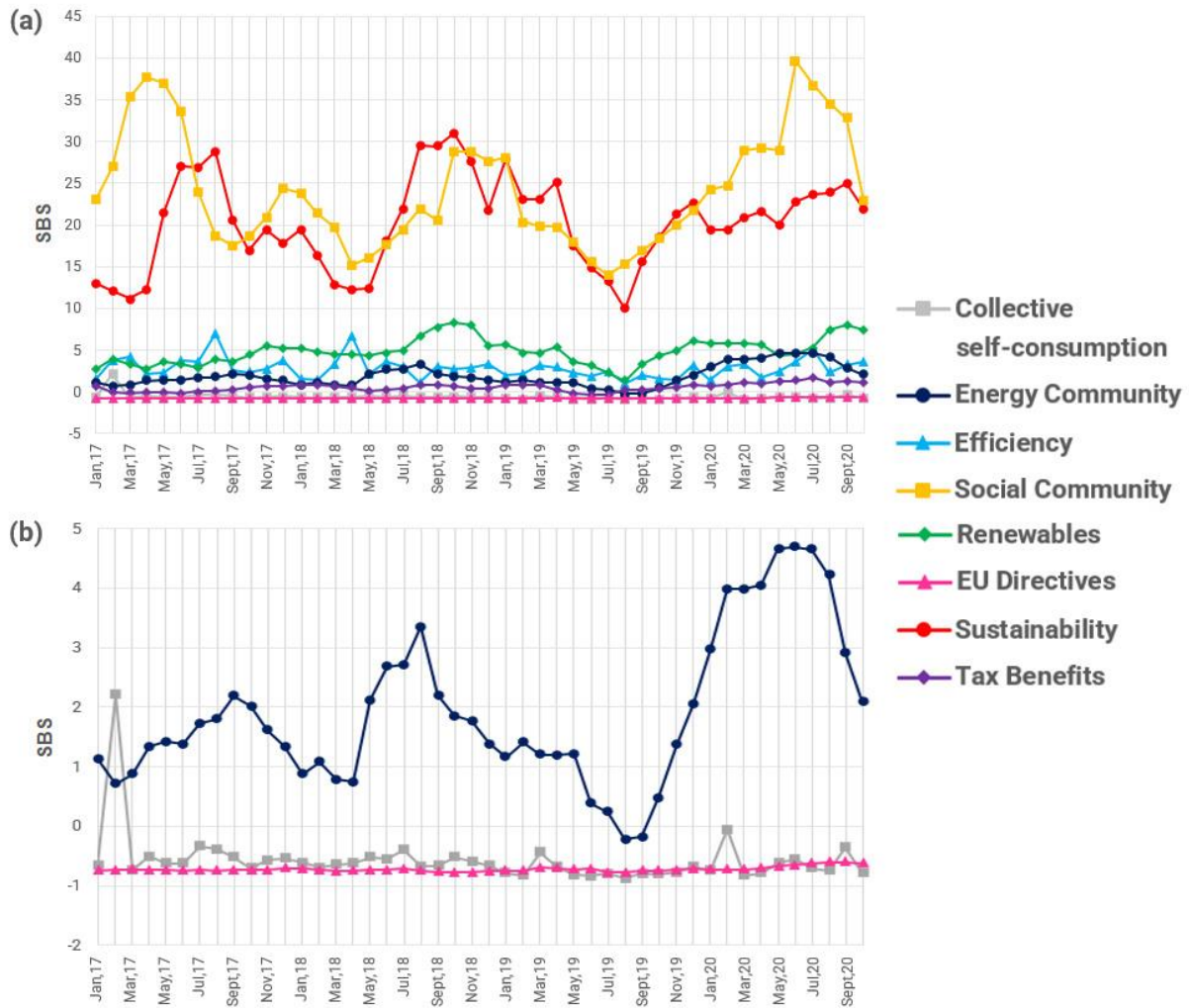

**Figure 2**. SBS time trends.

Figure 2b zooms on 3 ECKs: *energy community*, *collective self-consumption*, and *EU directives*. The SBS values for *EU directives* and *collective self-consumption* are negative for almost the entire period. This signals that these two topics were relatively unimportant in the energy-related online news of the past four years, being debated less than the average terms and topics. Concerning the SBS of *energy community*, we notice a dip around the middle of 2019, with a subsequent positive trend in 2020. This improvement in this topic's societal uptake is related to the publication of the Italian "Milleproroghe" decree n. 162, of December 30, 2019, and, thereafter, of the law n. 8, of February 28, 2020, enabling the establishment of energy communities in Italy. Reading the articles of 2020 related to *energy communities* allowed the verification and validation of this result and confirmed the positive impact of the "Milleproroghe" decree. On the contrary, the importance of *collective self-consumption*, which is actually regulated by the same laws, seems not to be affected by these events.

Interestingly, the trends of the analyzed ECKs in Figure 2b are unaffected by the publication of the EU Directive 2018/2001/EU on December 11, 2018. In particular, the lexical environment of the three ECKs is not populated by words connected to any of the specific references to this Directive, unlike national regulations, which showed to play a key role in the national news. The fact that EU directives are not very much cited in energy-related national news, as demonstrated by the results of this analysis, discloses that important EU-level information and new legislative trends have a minor diffusion at a national level. This with the potential effect of delaying the whole energy transition, with new solutions that take up to 2-3 years to become sufficiently well known. This gap further makes Italy a laggard and disadvantaged country in the European energy transition panorama, particularly with regard to energy community initiatives. Indeed, up to a few years ago, Italian regulations did not allow any community contract for the sharing of renewable energy, differently from other European countries [78]. This delay also affected technology development and market uptake of energy communities in Italy. Nevertheless, the situation is rapidly changing and, nowadays, Italy is keeping up with European policies with a lower delay than other laggard countries, e.g., Spain.

When assessing the media importance of specific topics through the SBS indicator, looking at the three dimensions of prevalence, diversity, and connectivity represents a key action. Figure 3 presents the overall score of these three components for each ECK. Results confirm that *sustainability* and *social community* appear in the news more frequently, with comparable scores of prevalence and diversity. However, the connectivity of *social community* is significantly lower than the one of *sustainability*, meaning that this topic is less integrated into the general discourse. Connectivity is even lower for all the remaining ECKs, indicating the need for increased media coverage of energy communities, as well as of narratives that can make the theme central in energy news.

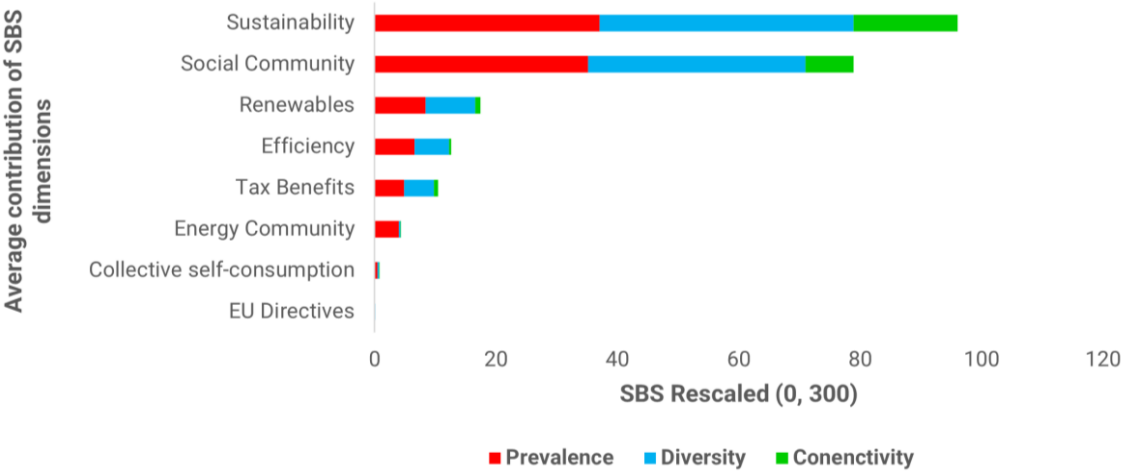

**Figure 3**. SBS dimensions.

Another interesting result comes from the analysis of ECK image similarity, based on the assessment of their textual associations. According to the idea that the more similar is the textual image of two ECKs, the closer they appear, Figure 4 shows that the image of *energy community* is associated with that of *sustainability* and *social community*. On the contrary, it is distant from *EU directives* and *collective self-consumption*, meaning that these last two ECKs are not well integrated into the discourse on energy communities and shared renewable energy, despite their technical connection and complementarity. In addition, this graph confirms the previous finding that *EU directives* is farther away from every other ECKs considered in the analysis – possibly suggesting reduced exploitation of new EU policies at their early stage on a national basis.

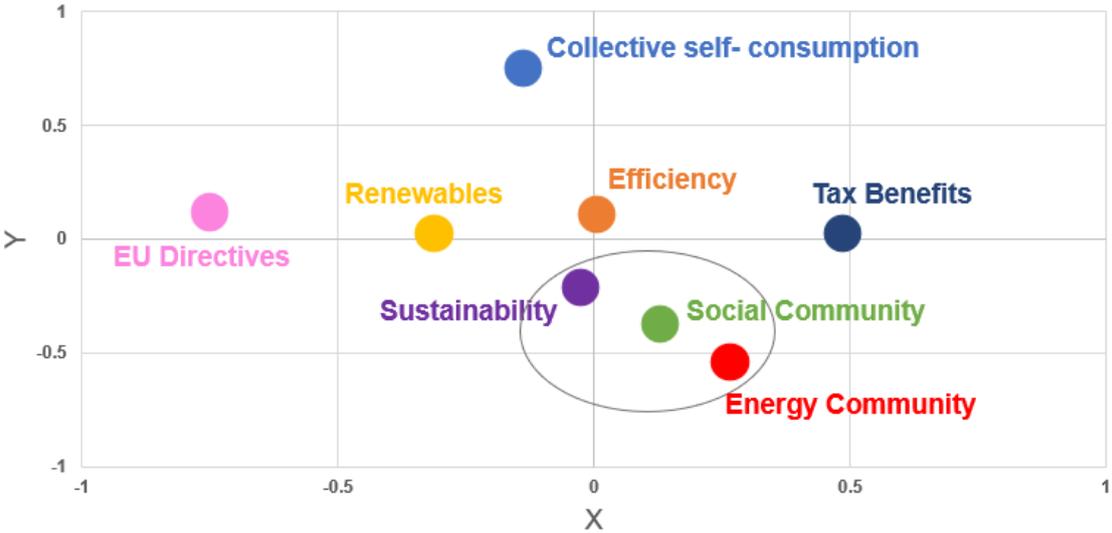

**Figure 4**. ECKs image similarity.

We additionally analyzed the most associated words with *energy community, sustainability,* and *social community*. The word clouds in Figure 5 provide a graphical representation of association frequencies. Figure 5a explores the main associations with *energy community*. There are numerous words with a positive valence, e.g., free, commercial, country, state, agreement, project, new, growth, and energy. However, the ECK seems mainly associated with words from an economic discourse, with weaker links to terms highlighting environmental and social benefits. The discourse around *energy community* seems primarily focused on the definition of the economic and institutional context enabling the implementation of this solution, somehow also detaching the discourse from the social dimension of the communities, which has been shown to represent another key driver for people engagement. Indeed, Italian news does not give evidence of the positive social implications of energy communities, which are considered particularly relevant at a European level. *Energy community* is also significantly associated with *sustainability* and *social community*, as pointed out previously. There are words that appear in all three word clouds: state, project, sector, economic, energy, region, and territory. Therefore, the image of the ECKs is similar when referring to *energy* concepts that have *economic* implications and concern the local and national *territory*. The strong association to the concepts of *regional* and *project* also shows that a relevant number of news cover specific initiatives while maintaining a local focus and attaching less importance to the role of energy communities as a key driver for sustainability and energy transition, as advocated in EU directives. This is also proved by the word *Florence,* referred to news reporting the initiatives supported by a local politician. Indeed, the actions taken by politicians and public figures become particularly relevant as they get significant media attention in Italy. Accordingly, leveraging these people's media exposure could be considered an effective strategy to promote *energy communities* implementation or, more in general, societal acceptance of any energy- and environment-related topic. Consistently, we also notice four unexpected words standing out in the *sustainability* word cloud that form the expression "Council member Federica Fratoni", the Councilor for the Environment and Soil Defense of Tuscany region in Italy. This result shows that the actions of public figures get relevant media attention in the energy-related discourse – thus potentially suggesting that a communication message conveyed by politicians could boost awareness of new energy solutions more than complex technical discussions. However, this strong link to aspects of sustainability was found only for a local politician and not with those known at the national level, thus demonstrating the weak importance attributed to this topic.

*Sustainability* (Figure 5b) is mostly linked with the energy and environmental discourse. Words such as retrofit, reduction, gas, zero, $CO_2$ are present, as well as other climate change-related words, e.g., greenhouse, change, and climate. Being uniquely associated with sustainability, these words could be among the causes of the high values assumed by its SBS. Indeed, global warming, climate change, and energy and environmental aspects have been much debated in recent years.

The last word cloud (Figure 5c) represents the image of *social community*. As also observed for *sustainability*, this topic is much better associated with political topics than *energy community*. In particular, it includes words referring to international (state, European), national (country, Italy), and local (region, Tuscany, Florence) entities and to projects and plans for economic development and innovation. The presence of words related to local entities and stakeholders suggests a strong association of the *social community* concept with strategies and policies meant to support innovation and economic growth at a local level.

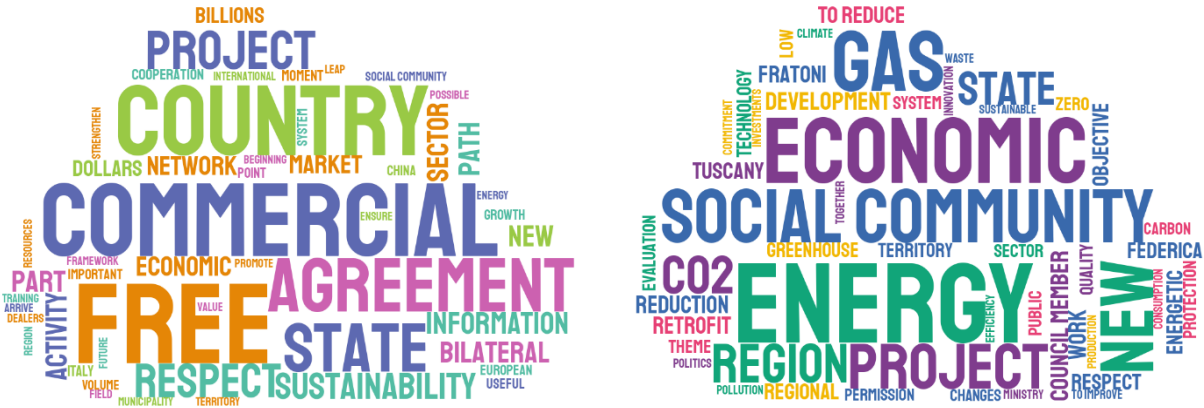

(a)                                                              (b)

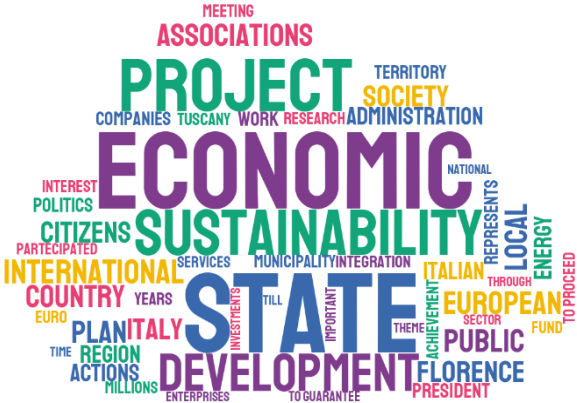

(c)

**Figure 5**. Main associations with a) *energy community*, b) *sustainability*, and c) *social community*.

## 5. Discussion and Conclusions

Within the context of the low-carbon energy transition, this work aims at evaluating the media debate on energy communities, potentially impacting societal awareness and citizen behavior. To this aim, we analyzed online news published in four years, up to the end of 2020. The approach we present allows continuous monitoring of online news and other potential textual data sources, such as social media, providing insights useful to support the energy transition. This has the advantage of having easily updatable results that could complement the findings of general surveys and specific interviews (which may only partially represent the societal awareness about a topic). To the best of our knowledge, this is the first scientific contribution that assesses the media importance – not just media coverage – of energy communities, and their related energy transition topics, using the SBS approach. In this work, we presented methods and tools useful for scholars, which can also provide immediate insights to practitioners and policymakers to increase citizens' awareness of energy communities and their importance for the energy transition. These insights are obtained from studying ECKs media importance over time and the analysis of textual associations. Key takeaways are summarized in Figure 6 and discussed in the following.

In general, we find that *social community* and *sustainability* topics are the most important during the whole period under analysis. This result is due to the broader scope of these two ECKs and consistent with existing research that found a deep connection between the concepts of sustainable cities and social communities [44]. By contrast, the topics of *energy community* and *collective self-consumption* have low importance values, with the former being more relevant than the latter. Despite their key technical and economic relevance, these two ECKs are not particularly debated in online news, ultimately leading to reduced diffusion and understanding among Italian stakeholders, as indicated by existing studies stressing the influence of media on citizens' perception of energy-related concepts [38,40]. Moreover, we find that the textual image of *energy community* is distant from that of *collective self-consumption*. In particular, the important aspects of self-consumption, self-production, and collective use of energy seem not well integrated into the energy communities discourse – despite the fact that energy communities consist of collective self-consumption of potentially green energy on a local basis. Therefore, the more effective integration of this topic into discourses related to energy communities and shared renewable energy could help explain the potential of this solution and make society better aware of this opportunity to proactively contribute to the energy transition through a bottom-up approach.

The media importance of *energy community* increased in 2020, showing a positive effect of the approval of a national law enabling the establishment of renewable energy sharing schemes in Italy. While this shows that the media cover national laws, it also proves that such coverage is not yet enough to boost the societal awareness of energy communities. Even worse, we found minimal media impact from the issuance of the EU Directive introducing energy communities. The national context seems disconnected from the European one, despite the EU directives representing the key policy reference for developing national laws and incentives. Indeed, words related to *EU directives* are of very low importance in Italian online news. This outcome sheds light on Italy's difficulty in promptly keeping up with European policies in the adoption of energy-efficient solutions, with non-negligible political, societal, and economic competitiveness penalties. This finding is consistent with a review of the community energy sector in Italy [14], which showed that a few existing Italian energy communities were implemented via ad-hoc models that work well on a local scale but are hard to replicate. In addition, as we find that politicians get particular media attention in Italy, the buzz generated by their actions could be leveraged to increase awareness of new energy solutions (as also highlighted by Azarova et al. [32]). However, no charismatic person at the national level, supporting energy communities, emerged from our study. Conversely, energy and sustainable development topics were strongly linked to virtuous local politicians. This finding highlights the pivotal role of public figures and local authorities in triggering and supporting community energy initiatives. Accordingly, Herbes et al. [17] demonstrated that the participation of municipalities or other local entities as members of these initiatives could push their societal uptake. In this view, the actions taken by public figures, as well as their communication, could be used to reach a wide general public and push citizens toward engaging in the energy transition. Indeed, our results regarding SBS of *energy community*, *collective self-consumption*, and *EU directives* show that energy laws and related energy strategies go unnoticed, if inaccurately addressed and communicated.

On the other hand, the findings of our work show a link between the topic of energy communities and those of sustainability and social communities. Understanding the similarity of these three topics is a key challenge since *sustainability* and *social community* are the ECKs with the highest SBS and among the concepts that could better support the energy transition [46] and, therefore, the societal uptake of energy communities. In particular, we find that the image of these three ECKs is similar in terms of reference to energy concepts that have economic implications and concern the territory. This result confirms the particular relevance of market and institutional discourses in the news involving the analyzed ECKs. Moreover, *social community* concept is shown to be strongly associated with policies supporting innovation and economic growth at a local level. Consistently and aligned with what has been shown in the work of Lyytimäki [41], we maintain that the awareness of energy communities could be improved if they were discussed more about innovation and economic growth. Indeed, the results show the focus of discussions around *energy community* on economic aspects, which are a key element to consider for their societal uptake. The economic implications of energy communities are often associated to the viability of renewable energy and related investments; this link is stressed in literature for a few other European countries [16] but not for all [79]. Therefore, the improvement of social acceptance of energy communities in Italy would benefit from communication campaigns clarifying economic aspects, such as the return on investment and long-term economic benefits, as well as realistic risks and complexities. Another way to improve the development of new energy communities could be to communicate their environmental and social benefits more strongly, thus improving people's interest and engagement. Lagendijk et al. [16] stressed a lack of connection in the media discourse of material and social aspects of the renewable energy transition and recommended energy communities be depicted as "beacons of transformation". Our results also stress the importance of communicating the link between technical, economical, and social benefits of energy communities. Past research has shown that wisely increasing the keywords related to a topic can help make it grow over time [45]. There are themes that, if properly stressed and linked with energy communities, could foster their development, such as highlighting not only technical and economical but also social (association with social communities) and environmental (association with sustainability) benefits – which are particularly disregarded in Italy, at the moment. Shedding light on the multiple benefits of energy communities and the connection between energy and social sustainability in communities would improve their media importance, ultimately affecting societal awareness. Indeed, every energy community has a social community at its base. This seems like an obvious but key insight highlighted by our analysis. Although our study refers to a specific case, this finding could be exploited in broader contexts, given the internationally acknowledged primary social purpose of energy communities towards collective fair energy "prosumption". Local renewable energy initiatives have received high positive consideration from the media in countries such as Denmark, the Netherlands, and Sweden [79], – which are pioneers in the implementation of energy communities. In particular, the democratic character involving citizens' cooperation is appreciated by the media in these countries. Indeed, Magnusson et al. [79] show a prevalence of social frames in the media reporting about renewable energy innovations, which are generally presented in positive terms. Instead, in Italian online news, we find that more attention is paid to the economic implications of renewable energy communities rather than to their social aspects. Lastly, Bauwens et al. [21] showed that the meaning of "community" is evolving and "there is a shift away from community as a process and an increasing emphasis on community as a place". Our results support this statement and further highlight the link of a community not only with a place, but also with the people living in that place, having their own social identity and energy signature.

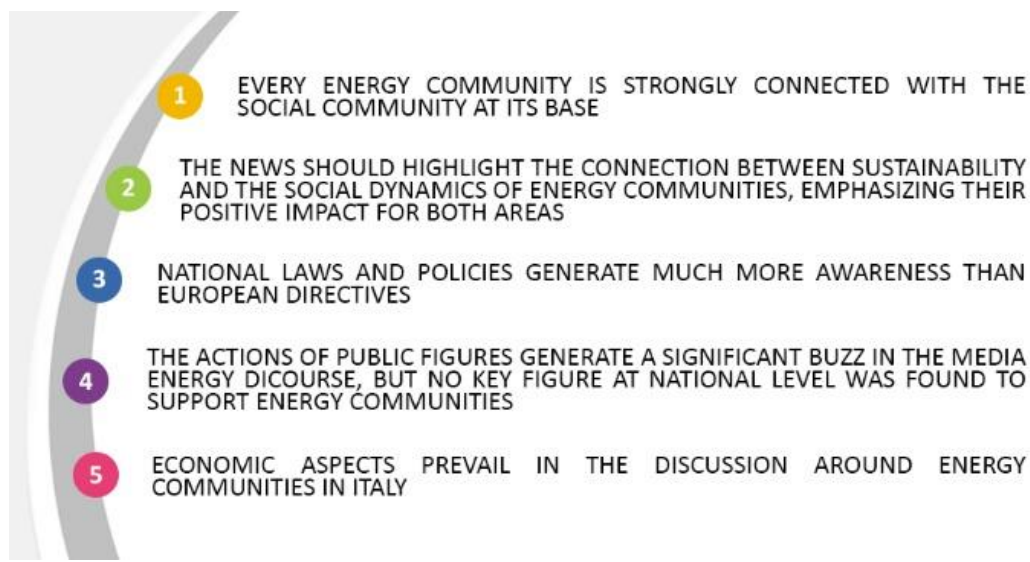

**Figure 6**. Main findings and insights from our case study.

Our study is limited to the Italian context, as a key example of a laggard country in the adoption of energy communities. Our analysis is limited to online news, with data until October 2020. Consequently, we ignore recent developments achieved during 2021, but we do this on purpose, as the COVID pandemic had a substantial impact on news and citizen behavior, and we do not want to bias this analysis. However, more recent news could be considered in future research. Future developments of this research may extend the analyzed database also in terms of sources by including, for example, the analysis of radio and TV broadcasts, printed news, social media (e.g., Twitter, Facebook, etc.), or specialized forums. Future work could also consider mixed methods approaches, where semantic network analysis is combined with interviews and workshops, and applied to the answers of citizens, practitioners, or experts. Community energy experts would represent a stakeholder other than online media, with a potentially different point of view. Comparing the views of different stakeholders could be very much interesting and deserves dedicated future research. Likewise, the SBS could be used to reveal key concepts in laws, regulations, and policies or in specific energy community text documents. It could be important to combine the information from our approach with information from other methodologies (e.g., survey-based), not focusing on just one source of data, such as web-only, policy-only, and interview-only.

In conclusion, this study uses a novel methodology to evaluate the media discourse in Italy and identifies barriers and opportunities for the diffusion and social acceptance of energy communities. Our approach allows continuous monitoring of media importance trends of energy transition topics to obtain insights for developing strategies at the national and local level – potentially exploitable for achieving societal, environmental, and economic benefits. Our analysis highlights communication gaps and helps define strategies to increase the social awareness of energy-related topics, starting, for example, from a better communication of EU directives. The societal ground should be prepared to embrace the use of new energy transition solutions, with key benefits in terms of people engagement and economic-energy competitiveness.

**Acknowledgments**

The authors' acknowledgments are due to the European Union's Horizon 2020 program under grant agreement No 890345 (NRG2peers). The authors are grateful to Vincenzo D'Innella Capano, CEO of Telpress International B.V., and to Lamberto Celommi, for making the news data available and for the support received during the data collection process. The computing resources and the related technical support used for this work were provided by CRESCO/ENEAGRID High Performance Computing infrastructure and its staff. CRESCO/ENEAGRID High Performance Computing infrastructure is funded by ENEA, the Italian National Agency for New Technologies, Energy and Sustainable Economic Development and by Italian and European research programs. This work was also partially supported by the University of Perugia – Department of Engineering, through the program "Fondo Ricerca di Base 2021", project n. RICBA21AFC ("Business Intelligence, data analytics e simulazioni ad eventi discreti per l'ottimizzazione di decisioni e performance aziendali, nell'era dell'Industria 4.0"). The funders had no role in study design, data analysis, decision to publish, or preparation of the manuscript.